# Coarse-Grained Molecular Dynamics Simulations for Oxidative Aging of Polymers under Various $O_2$ Concentrations

*Takato Ishida*, Kazuya Haremaki, Yusuke Koide, Takashi Uneyama, Yuichi Masubuchi*

Department of Materials Physics, Nagoya University, Furo-cho, Chikusa, Nagoya 464-8603, Japan

* E-mail: ishida@mp.pse.nagoya-u.ac.jp

## ABSTRACT

Modeling of polymer oxidative aging has been actively studied since the 1990s. Insights from these studies suggest that the transport of oxygen and radicals significantly influences aging heterogeneity, alongside chemical reaction kinetics. A recent simulation study [Ishida et al., *Macromolecules*, 56(21), 8474-8483, 2023] demonstrated that mesoscale heterogeneity arises when the H-abstraction reaction occurs faster than the relaxation times of polymer chains. In this study, the simulations were extended by modeling the rate of oxygen addition to polymer radicals ($k_2$) to reflect the effects of the $O_2$ concentration. Three key aspects of oxidative aging behavior were found to be influenced by the $O_2$ addition rate: (i) reaction kinetics, (ii) the degree of heterogeneity, and (iii) amount of crosslinking. Namely, reducing $O_2$ concentration slows the conversion of polymer radicals into H-abstractable peroxyl radicals. This deceleration delays H-abstraction reactions, increases the number of polymer radicals, and promotes crosslinking reactions between two polymer radicals.

## 1. INTRODUCTION

There is no doubt that $O_2$ concentration plays a crucial role in polymer oxidative aging, as confirmed by several experimental works [1–4]. The significance of $O_2$ originates from the fact that the reaction mechanism governing oxidative aging is represented by the "closed-loop mechanistic scheme" (CLMS) [5–10] based on autoxidation reactions [11,12]. Numerous studies have successfully described the kinetics of oxidative aging reactions through kinetic modeling based on the CLMS [5–10]. This framework has significantly advanced the modeling of aging processes at the microscopic reaction level. The reaction rate of $O_2$ addition to a polymer radical (P·), a key parameter in the CLMS, is proportional to $O_2$ concentration, reflecting the reaction rate constant for $O_2$ addition (so-called $k_2$) [11].

Another critical aspect is the diffusion-limited oxidation (DLO) effect [13], which induces macroscopic heterogeneity in aging behavior along the thickness direction of the sample. DLO arises when oxygen consumption exceeds the $O_2$ supply, lowering the internal $O_2$ concentration relative to the surrounding environment. DLO systems can be numerically analyzed using a continuum field framework that accounts for material transport and reaction kinetics, with previous studies reporting simulations based on the finite element method (FEM) [10,13]. In such approaches, the dynamics of polymers are not explicitly considered. However, aging reactions are intrinsically linked to the transport of macroradicals governed by polymer chain dynamics [14]. Thus, DLO simulations should incorporate a heterogeneous $O_2$ concentration field into a model capable of considering polymer dynamics, performed on a system size large enough to observe DLO effects (typically at the sub-mm scale [10]). There are two approaches to appropriately handle the above considerations. One is to simulate a system with a very large size and explicitly solve the $O_2$ molecule dynamics or the $O_2$ concentration field. The other is to employ a relatively small system size, assuming the $O_2$ concentration as a homogeneous field, and combining it with a conventional continuum field framework for scaling up. The former approach is impractical as it requires computational resources on an unfeasibly large scale. In contrast, the latter approach allows the effects of $O_2$ concentration to be represented by modifying the reaction rate of the $O_2$ addition reaction in the CLMS, making simulations feasible with realistic computational resources. Therefore, it is valuable to deepen our insights into oxidative aging simulations in mesoscale systems with varying $O_2$ concentrations.

Recently, we developed a coarse-grained molecular dynamics simulation (CGMD) capable of appropriately handling oxidative aging, including polymer segment and radical dynamics. In our previous studies [14,15], for simplicity, we performed simulations under oxygen excess conditions, known as the OER (Oxygen Excess Regime). As mentioned above, it is crucial to develop an aging simulation framework capable of accommodating changes in $O_2$ concentration. This study updates the aging simulation by examining the effects of $O_2$ concentration through variations in the $O_2$ addition rate to polymer radicals, focusing on its impact on reaction kinetics and aging heterogeneity.

## 2. MODELS AND SIMULATIONS

In this study, molecular dynamics calculations follow a framework of previous works [14,15]. To incorporate $O_2$ concentration dependence, only the set of chemical reactions (oxidation) and reaction kinetic parameters are modified from those in prior works [14,15]. In this model, polymer chains are represented using the Kremer-Grest (KG) model [16], with the total potential determined by the sum of the finite extensible nonlinear elastic (FENE) spring potential ($U_{\mathrm{FENE}}$) between chemically bonded beads, and the Weeks-Chandler-Andersen potential ($U_{\mathrm{WCA}}$) acting between all particles. $U_{\mathrm{FENE}}$ and $U_{\mathrm{WCA}}$ are expressed as follows.

$$U_{\mathrm{FENE}}(r) = \begin{cases} -0.5kR_0^2 \ln\left[1 - \left(\frac{r}{R_0}\right)\right], & r \le R_0 \\ \infty, & r > R_0 \end{cases} \tag{1}$$

$$U_{\mathrm{WCA}}(r) = \begin{cases} 4\varepsilon\left[\left(\frac{\sigma}{r}\right)^{12} - \left(\frac{\sigma}{r}\right)^{6} + \frac{1}{4}\right], & r \le 2^{1/6}\sigma \\ 0, & r > 2^{1/6}\sigma \end{cases} \tag{2}$$

In these equations, $r$ is the distance between interacting or chemically bonded beads, $\varepsilon$ is the WCA energy parameter, and $\sigma$ represents the bead diameter. $k$ is the FENE spring constant and $R_0$ corresponds to the maximum bond length. These parameters are chosen as the same as the standard settings of the KG model [16], being $R_0 = 1.5\sigma, k = 30\varepsilon/\sigma,$ and $\rho^* = 0.85/\sigma^3$. The position of each bead obeys the Langevin equation, which describes the balance among inertia, conservative, drag, and random forces, as written below.

$$m\frac{d^2\boldsymbol{r}_i}{dt^2} = -\frac{\partial U_{\text{tot}}}{\partial \boldsymbol{r}_i} - \Gamma\frac{d\boldsymbol{r}_i}{dt} + \boldsymbol{W}_i(t) \tag{3}$$

where $m$ is the mass of the bead, and $\boldsymbol{r}_i$ is the position of the $i$-th bead. $\Gamma$ is the bead friction coefficient, set to 1.0 in this study. $\boldsymbol{W}_i(t)$ represents Gaussian random force that satisfies $\langle \boldsymbol{W}_i(t)\rangle = 0$ and $\langle \boldsymbol{W}_i(t)\boldsymbol{W}_j(t')\rangle = 2k_BT\Gamma\delta_{ij}\delta(t-t')\mathbf{I}$. $k_B$ is the Boltzmann constant and $T$ is the temperature. Here, $\langle\cdots\rangle$ indicates the statistical average and **I** is the unit tensor. $U_{\text{tot}}$ is the total potential of the system written as:

$$U_{\text{tot}} = \sum_{i>j} U_{\text{WCA}}(r_{ij}) + \sum_{i,j\in S_{\text{bond}}} U_{\text{FENE}}(r_{ij}) \tag{4}$$

where $r_{ij}$ is defined as $r_{ij} = |\boldsymbol{r}_i - \boldsymbol{r}_j|$, and $S_{\text{bond}}$ represents the set of bonded pairs of beads.

In this work, the bead mass is set to $m = 1$, with the units of length, energy, and time defined as $\sigma$, $\varepsilon$, and $\tau = \sigma\sqrt{m/\varepsilon}$, respectively. The temperature is set to $k_BT = 1$. Here, a cubic simulation box with a side length 67.2 is considered under periodic boundary conditions. A polymer melt system consisting of 2560 Kremer-Grest chains with a polymerization degree of $N = 100$ is placed in the simulation box. Numerical integration of the Langevin equation is performed with a time step of $\Delta t = 0.005$, and all simulations in this study are conducted in the *NVT* ensemble.

The oxidation reactions in this simulation are based on a CLMS-derived set of chemical reactions [2,6,17]. Except for the inclusion of an additional $O_2$ addition reaction to polymer radicals, the reaction mechanisms considered are consistent with previously reported chemical reactions for the thermal oxidative aging of polypropylene (PP) [14,15].

$$\mathrm{POOH} \rightarrow \mathrm{PO}\cdot + \cdot\mathrm{OH} \qquad k_{1u} \qquad (5)$$

$$2\mathrm{POOH} \rightarrow \mathrm{PO}\cdot + \mathrm{POO}\cdot + \mathrm{H_2O} \qquad k_{1b} \qquad (6)$$

$$\mathrm{P}\cdot + \mathrm{O_2} \rightarrow \mathrm{POO}\cdot \qquad k_2 \qquad (7)$$

$$\mathrm{PH} + \mathrm{POO}\cdot \rightarrow \mathrm{P}\cdot + \mathrm{POOH} \qquad k_3 \qquad (8)$$

$$\mathrm{PH} + \cdot\mathrm{OH} \rightarrow \mathrm{P}\cdot + \mathrm{H_2O} \qquad k_3 \qquad (9)$$

$$\mathrm{PO}\cdot \rightarrow \mathrm{P}\cdot(\mathrm{end}) + (\mathrm{scission}) \qquad - \qquad (10)$$

$$2\mathrm{P}\cdot \rightarrow (\mathrm{crosslink}) \qquad - \qquad (11)$$

$$\mathrm{P}\cdot + \mathrm{P}\cdot(\mathrm{end}) \rightarrow (\mathrm{crosslink}) \qquad - \qquad (12)$$

Here, P· denotes a polymer radical, PO· represents an alkoxy radical, and POO· is a peroxy radical, all of which are types of macroradicals attached to polymer chains (Figure 1). POOH corresponds to a hydroperoxide group on polymer chains, while ·OH represents a free radical.

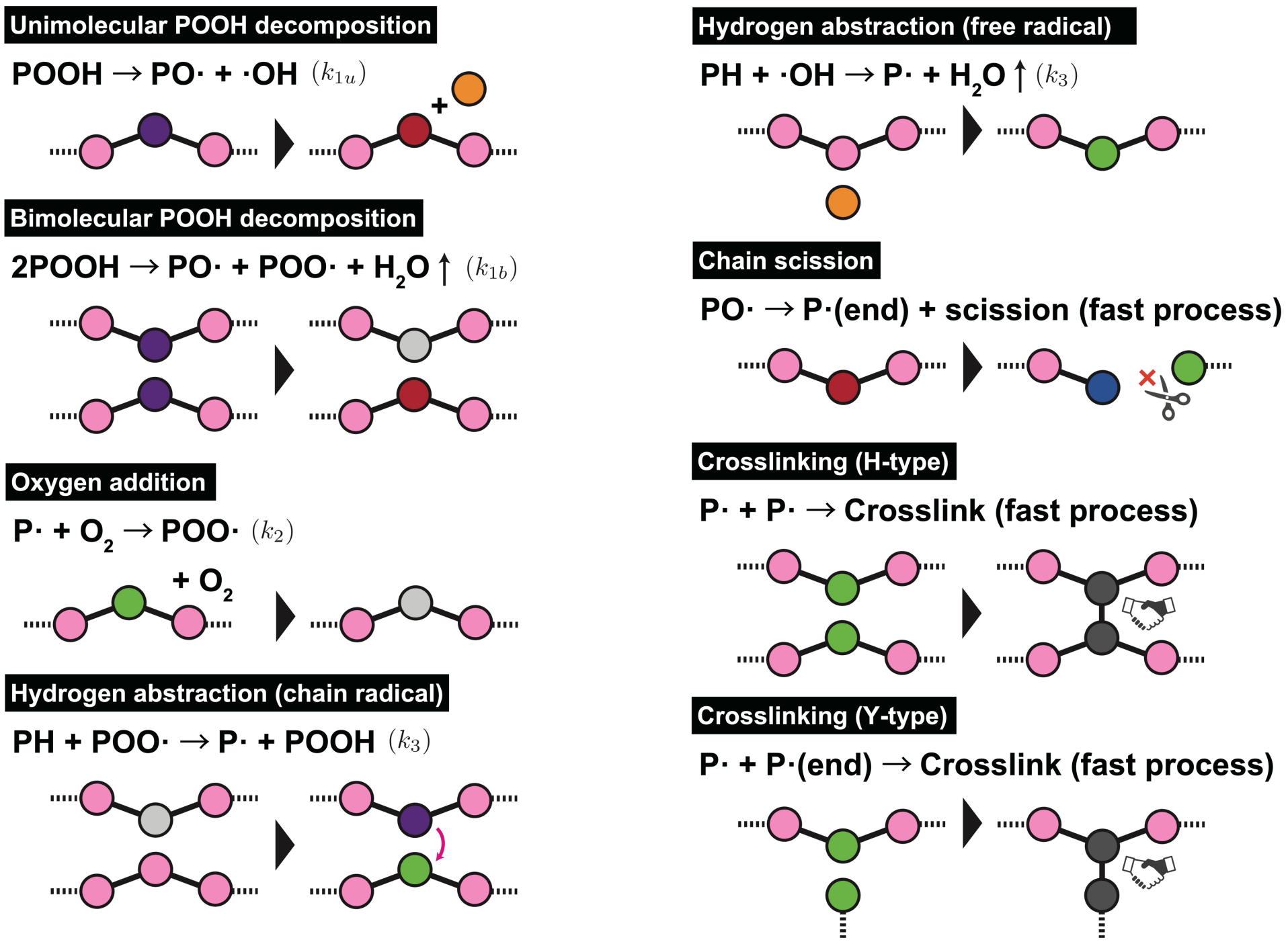

Figure 1 Schematic representation of oxidation reaction topology. PH (polymer substrate): pink, POOH (hydroperoxide): purple, PO· (alkoxy radical): red, ·OH (free radical): orange, P· (polymer radical): green, POO· (peroxy radical): white, scission end: blue, crosslinked beads: gray.

Among the chemical reactions considered, the low-molecular-weight species excluding polymer segments are $O_2$, OH·, and $H_2O$. In this work, the dynamics of $O_2$ molecules are not explicitly solved. Consequently, the $O_2$ concentration, as a control parameter in this study, is introduced as a uniform $O_2$ concentration field in the system. Thus, $O_2$ addition reactions can be treated as first-order kinetics scaled by $O_2$ concentration. On the other hand, ·OH beads are modeled as being driven by the Langevin equation without interacting with other particles. Although the diffusivity of ·OH only depends on the friction it experiences from the surrounding medium, it is known that this friction coefficient has a limited impact on oxidative aging [15]. Thus, we assign ·OH beads a sufficiently low friction coefficient, enabling them to traverse distances beyond the system size within the average time ($\sim 1/k_3$) before deactivating through H-abstraction reaction. $H_2O$ molecules generated during oxidation are immediately removed from the system and are not included in the dynamics calculations.

Here, chemical reactions are modeled to occur stochastically when the reaction topology shown in Figure 1 is satisfied. The reaction cutoff distance is set to $r_{\mathrm{cutoff}} = 2^{1/6}$. The rates of considered reactions ($k_{1u}, k_{1b}, k_2, k_3$) are presented along with their corresponding reaction equations, assuming that the reaction topology is satisfied. For reactions labeled as "fast process" in Figure 1, where $k_{(\cdot)}$values are not specified, the reaction occurs immediately upon satisfying the reaction topology. The reaction kinetics are mapped onto the timescale of this simulation using parameters from previously reported thermal-oxidative aging of PP at 180 °C [2,6,17]. To investigate the effects of $O_2$ concentration on aging heterogeneity, the same parameter set as in prior studies [14,15] (with the exception of $k_2$)—where H-abstraction occurs before chain relaxation—was employed. If the reaction topology shown in Figure 1 is satisfied, the reaction rates expressed in the simulation time unit $\tau$ are as follows: $k_{1u} = 1.6 \times 10^{-5}/\tau$, $k_{1b} = 5.8 \times 10^{-4}/\tau$, $k_3 = 1.0 \times 10^{-1}/\tau$, and the initial (unaged) KG chain relaxation time $\tau_{R_0} = 2 \times 10^4 \tau$.

We perform aging simulations under various $O_2$ concentration conditions. Since the $O_2$ concentration in the system is proportional to the rate of the $O_2$ addition reaction $k_2$ [2], oxidative aging simulations are conducted for several $O_2$ concentrations by systematically varying $k_2$. Here, four cases of $k_2\tau$ are considered: $5.0 \times 10^{-4}$, $2.5 \times 10^{-3}$, $2.5 \times 10^{-2}$ and $5.0 \times 10^{-2}$. The maximum $k_2$ case is set to align with the kinetic model for the onset of OER in previous studies [2,6,17]. In OER case of $k_2\tau = 5.0 \times 10^{-2}$, the generated P· immediately undergoes $O_2$ addition and is converted to POO·. In non-OER cases, the probability of this reaction is adjusted based on $k_2$ to introduce a delay in the conversion of P· to POO·.

In this study, simulations were performed using Large-scale Atomic/Molecular Massively Parallel Simulation (LAMMPS), version 23Jun22 [18], with chemical reactions incorporated into the dynamics calculations via the REACTION package [19,20]. As in the previous works [14,15], the system was initialized from a fully relaxed state using a combination of the fast push-off and double-bridging methods. Initially, all beads were PH, but one bead was substituted with POOH, and the oxidative aging simulation started at $t = 0$ when this POOH bead decomposed.

## 3. RESULTS AND DISCUSSION

Figure 2 shows snapshots of oxidative aging simulations at different $k_2$, projected onto the $xy$-plane. Here, the conversion ratio $\alpha$ is defined as the residual fraction of PH beads, specifically $\alpha = 1 - \frac{N_{\mathrm{PH}}(t)}{N_{\mathrm{PH}_0}}$, where $N_{\mathrm{PH}}(t)$ is the number of PH beads at time $t$, and $N_{\mathrm{PH}_0}$ is the initial number. In Figure 2, only scission ends, P·, and POO· beads are visualized, and the bead size is displayed larger than the actual size ($1\sigma$) for clarity. The areas appearing as blank domains are actually filled with other types of beads, such as PH or POOH. The snapshots reveal that when $k_2\tau = 5.0 \times 10^{-2}$, the distribution of scission sites is highly heterogeneous. This phenomenon occurs because, as discussed in previous studies [14,15] under OER conditions, H-abstraction reaction by POO· radicals precedes chain relaxation. Conversely, as $k_2$ decreases, the number of P· radicals increases, and the spatial distribution of aged domains becomes more uniform. The average time for P· to convert to POO· (H-abstractable) via $O_2$ addition process is $1/k_2$. When $k_2$ is low, radicals remain in the P· state and tend to diffuse over longer distances. Even if the H-abstraction rate becomes larger than the average chain relaxation rate, the aging heterogeneity is somewhat mitigated due to decreased $O_2$ concentration. This characteristic aspect reflects the essential nature of $O_2$ concentration dependence in heterogeneous oxidative aging.

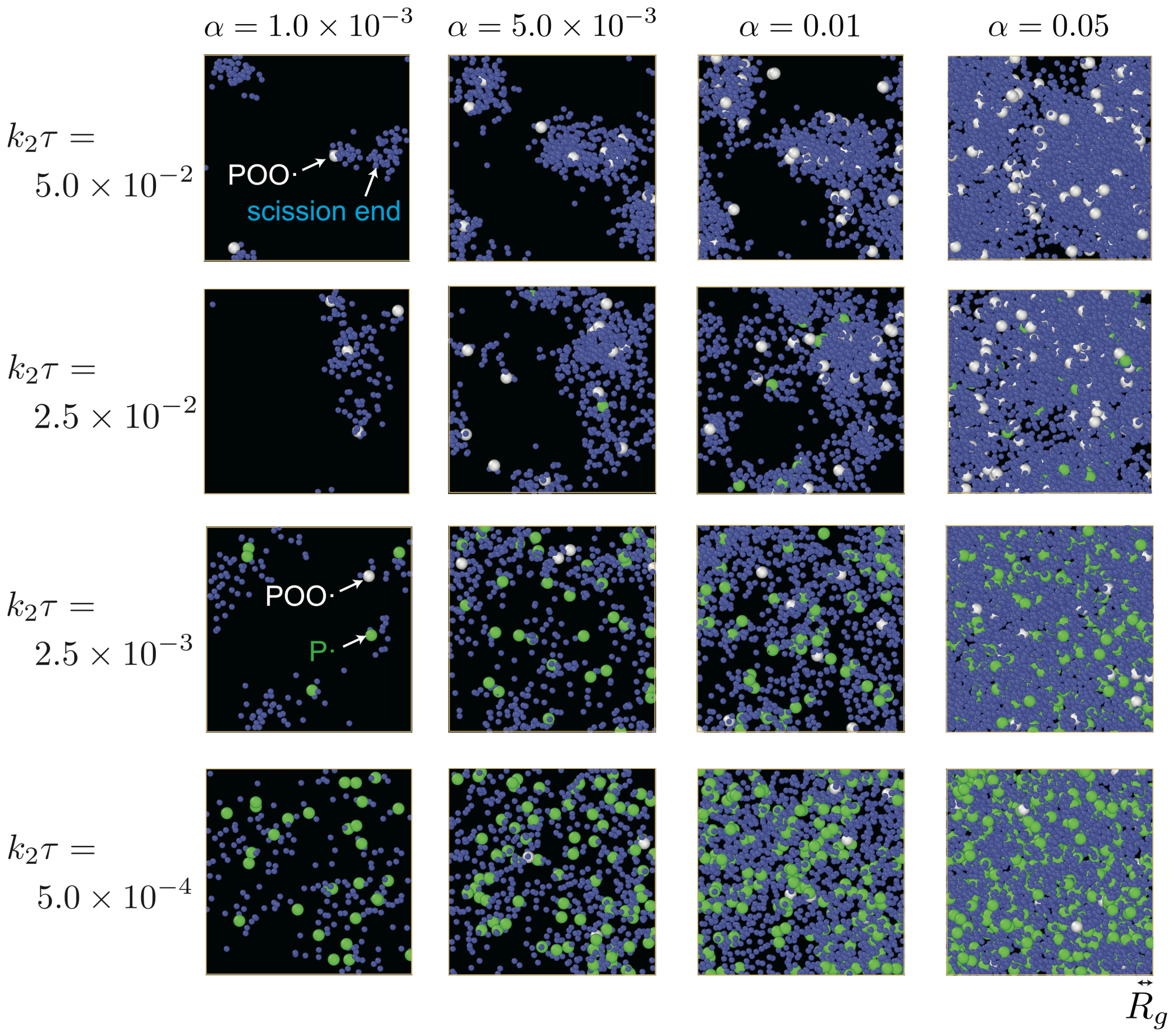

Figure 2 Snapshots from the oxidation aging simulations at various conversion levels $\alpha\left(= 1 - \frac{N_{\mathrm{PH}}(t)}{N_{\mathrm{PH}_0}}\right) = 1.0 \times 10^{-3}, 5.0 \times 10^{-3}, 0.01, 0.05$ under different $k_2$ conditions. Green beads represent P· (polymer radicals), white beads indicate POO· (peroxy radicals), and blue beads show scission sites. Black areas are filled with unreacted polymers.

Figure 3 presents the chemical reaction kinetics. Each plot exhibits autocatalytic behavior involved with an induction period, where the reaction accelerates as the concentration of the unstable product POOH (radical source) and the total amount of radicals increase. The kinetics exhibit a clear dependence on $O_2$ concentration, with lower $k_2$ leading to retardation of the oxidative aging kinetics. The $O_2$ concentration dependence of chemical kinetics is qualitatively consistent with experimental findings reported [1,2,21]. It is well known that the kinetics of oxidative aging kinetics are predominantly governed by the rate of H-abstraction reaction by POO· radicals [11]. In this study, we confirm that changes in $k_2$ alter the $O_2$ addition rate of P· → POO·, resulting in an increased ratio of P· to POO· at lower $O_2$ concentration conditions, as shown in Figure 4(a). Additionally, the rate of crosslinking reactions depends on the amount of P·. Although POO· may contribute to unstable crosslinks, such as POOP or POOOOP, their high instability renders them negligible in the examined conditions [11,22]. Consequently, with lower $k_2$ values, where P· is more prevalent, the ratio of crosslinking to chain scission ($N_{\mathrm{crosslink}}/N_{\mathrm{end}}$) increases, as shown in Figure 4(b). In contrast, almost no crosslinking occurred under OER conditions. In case of high-temperature thermo-oxidative aging, particularly near the sample surface where $O_2$ consumption is intense, the DLO effect often causes a decrease in the $O_2$ concentration within the bulk region of polymeric materials [13,23]. This effect is frequently discussed in macro-scale oxidative aging experiments [1,13,23–26]. Among these studies, experiments have reported that regions exhibiting the DLO effect tend to show a higher rate of crosslinking and a shift in the scission-to-crosslinking ratio toward crosslink dominance [24–26]. Our study successfully replicates these experimental observations.

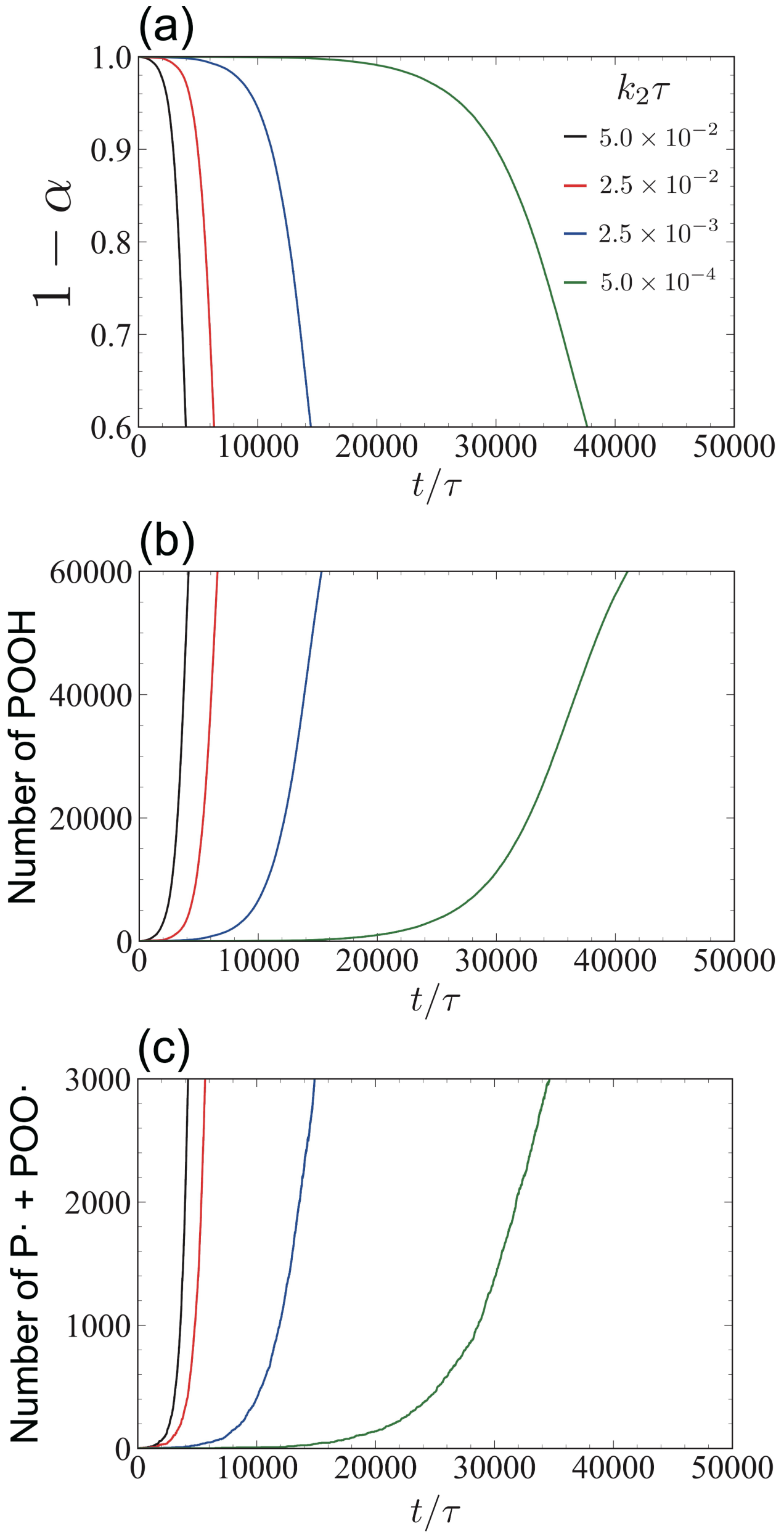

Figure 3 Time evolution of (a) the conversion $1-\alpha=\frac{N_{\mathrm{PH}}(t)}{N_{\mathrm{PH}_0}}$, (b) the number of POOH beads, and (c) the number of radical beads (P· + POO·).

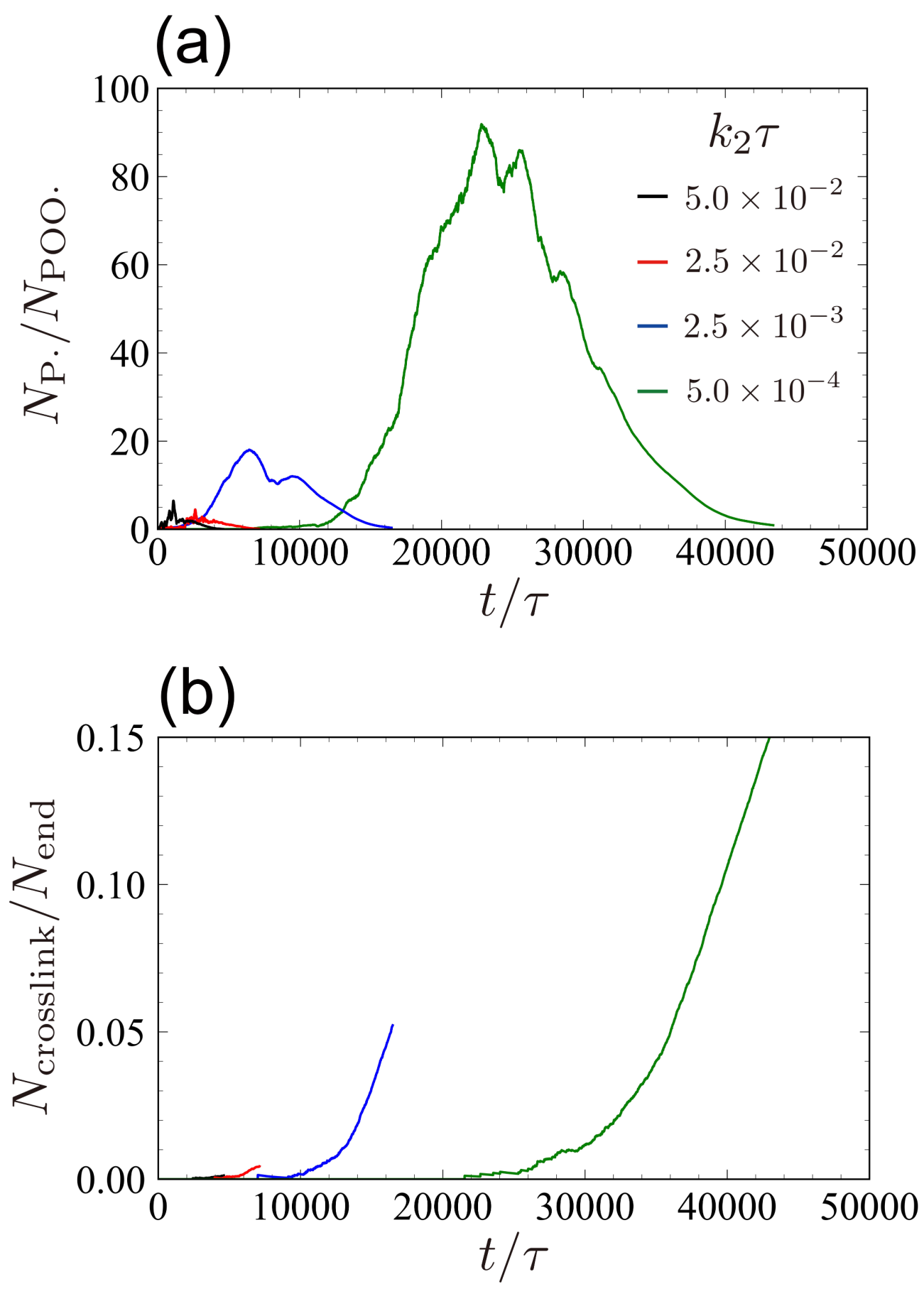

Figure 4 Time evolution of (a) the ratio of P· to POO· radicals ($N_{\mathrm{P}\cdot}/N_{\mathrm{POO}\cdot}$), and (b) the ratio of crosslink to scission end counts ($N_{\mathrm{crosslink}}/N_{\mathrm{end}}$) under various $k_2$ conditions.

Figure 5 illustrates the dependence of the induction period ($t_{\mathrm{ind}}$) and reaction time ($t_{\mathrm{react}}$) on $O_2$ addition rate. The induction period is defined according to our previous work [15], and the reaction time is defined as the duration from $t_{\mathrm{ind}}$ to reaching $\alpha = 0.5$. Both $t_{\mathrm{ind}}$ and $t_{\mathrm{react}}$ increase as $k_2$ decreases. They exhibit power-law behavior with respect to $O_2$ concentration, following the relations $t_{\mathrm{ind}} \propto {k_2}^{-0.45}$ and $t_{\mathrm{react}} \propto {k_2}^{-0.42}$, respectively. The impact of the $O_2$ concentration for the induction period in thermo-oxidative aging of PP has been experimentally demonstrated by Heude et al. [21]. Our simulations qualitatively capture this behavior. The aging kinetics of oxidative aging proceed through an accelerated period, and experimental evaluation of $t_{\mathrm{react}}$ for high $\alpha$ is challenging due to the volatilization of oxidized products and its uncertainty in molar extinction coefficients (in the case of infrared spectroscopy) [27]. Nevertheless, simulations provide a clear depiction of $k_2$ dependence of $t_{\mathrm{react}}$. The retardation of aging kinetics described above arises from differences in the H-abstraction reaction rate. The rate of this reaction is influenced by the number of POO· radicals and the spatial coordination of PH beads around POO· radicals. The latter contribution can be evaluated through the radial distribution function ($g_{\mathrm{PH-POO\cdot}}(r)$) between PH and POO·. Figure 6 presents $g_{\mathrm{PH-POO\cdot}}(r)$ at a conversion level of $\alpha = 0.1$ under different $k_2$ conditions. According to the results, there is no noticeable $k_2$ dependence in the local arrangement of PH around POO·. This indicates that the kinetics of oxidative aging are predominantly governed by the number of POO· radicals present in the system. In low $k_2$ cases, the reduced number of POO· radicals leads to a retardation of the overall chemical kinetics. Additionally, the power-law exponents of $t_{\mathrm{ind}}$ and $t_{\mathrm{react}}$ with respect to $k_2$ are both weaker than -1, indicating that the reaction kinetics are not dominated by a single reaction. One possible factor is that as $k_2$ decreases, the concentration of P·, the reactant in $O_2$ addition reaction, increases. Therefore, the overall reaction kinetics cannot be reduced to a single curve using $k_2 t$ (Figure S1 in Supporting Information). However, by normalizing with $t_{\mathrm{ind}}$, the results from simulations with different $k_2$ values are shown to align on a comparable timescale (Figure S2). Nonetheless, the clear $k_2$-dependence of the ratio of P· to POO· radicals and the ratio of crosslink to scission end counts is preserved (Figure S3, S4).

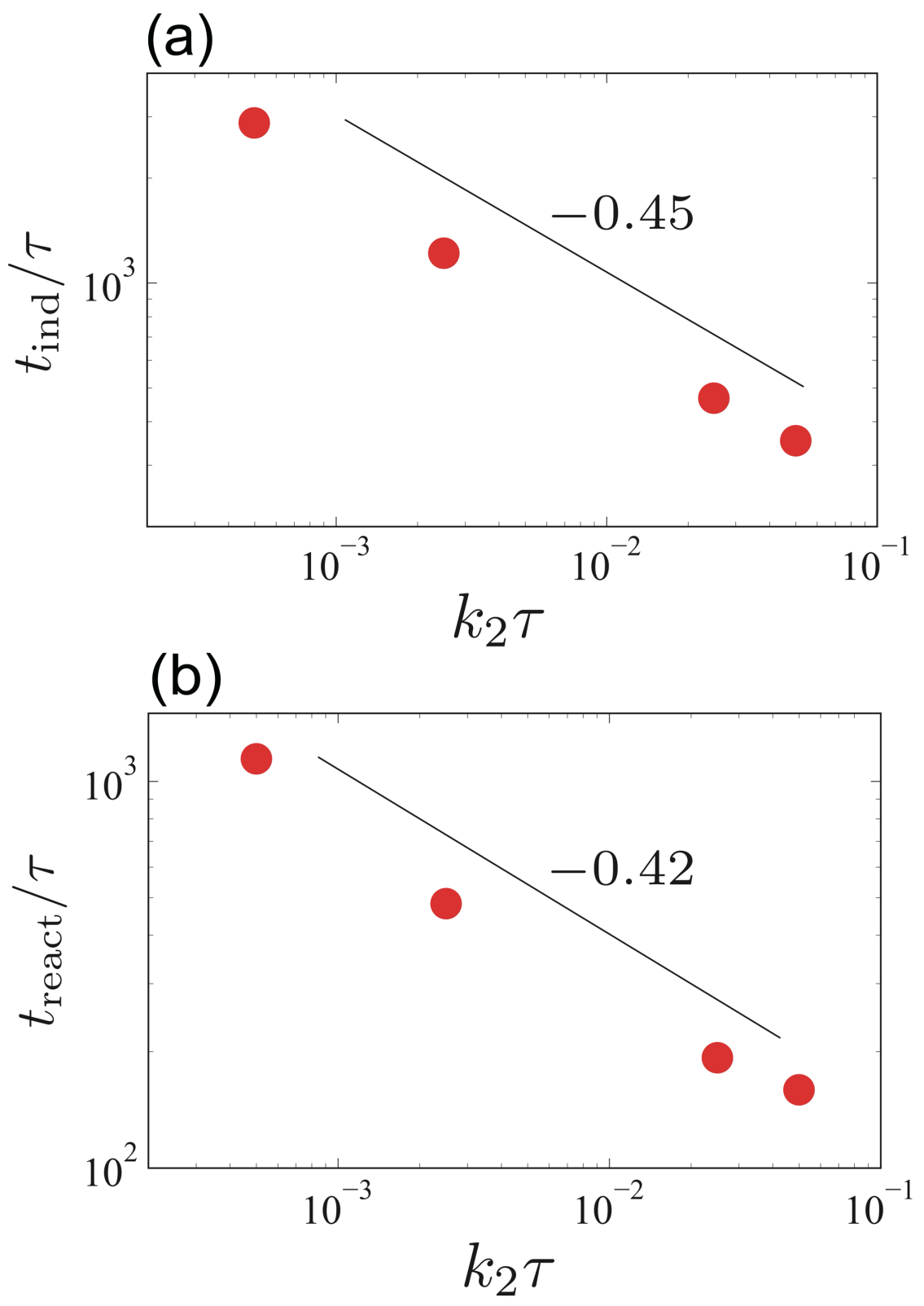

Figure 5 $O_2$ addition rate $k_2$ dependence of (a) $t_{\mathrm{ind}}$ and (b) $t_{\mathrm{react}}$. The figure displays the average results from 16 simulations initiated from different initial configurations and random seeds.

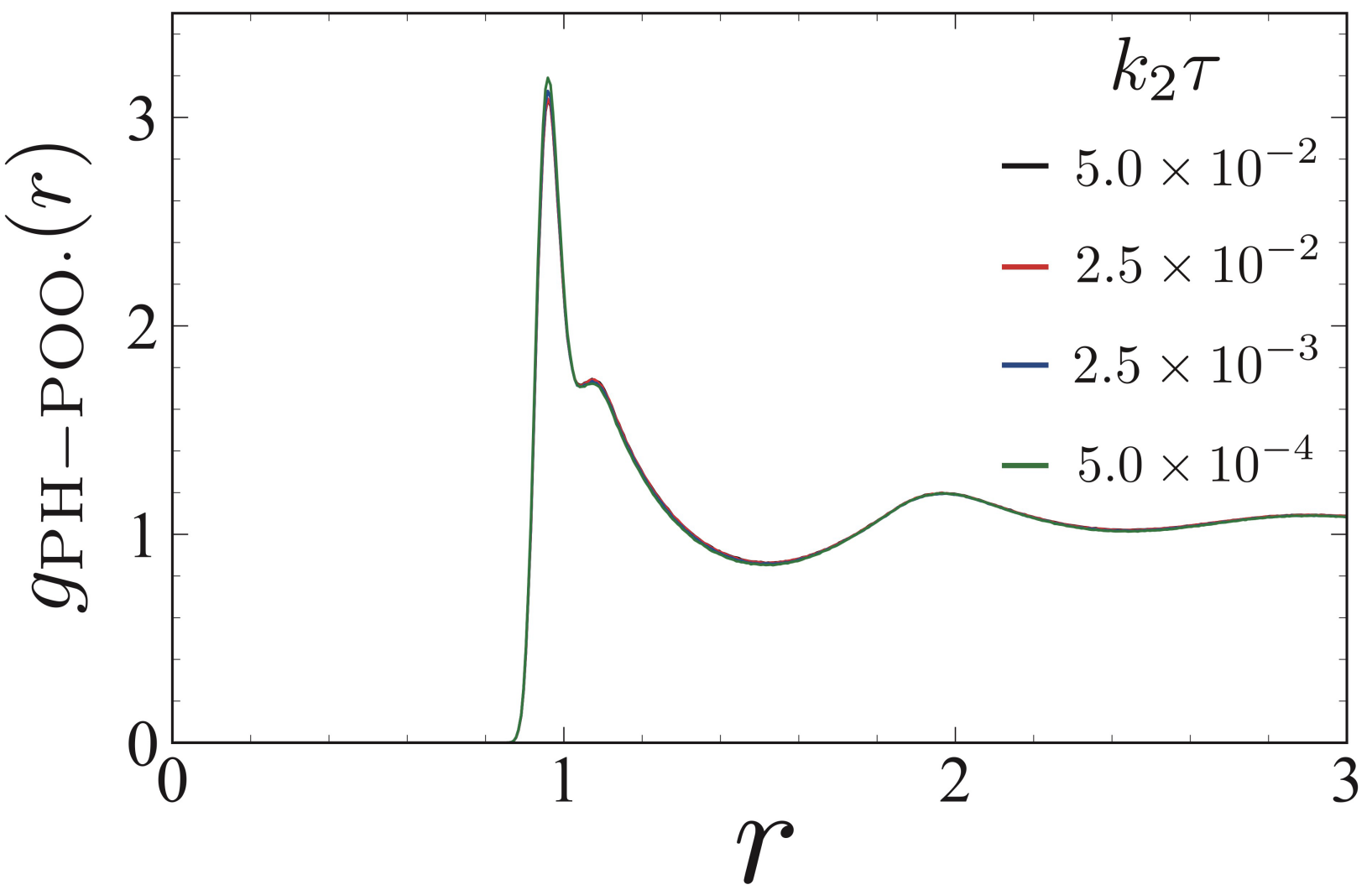

Figure 6 Radial distribution functions ($g_{\mathrm{PH-POO\cdot}}(r)$) between PH and POO· at $\alpha = 0.1$.

Next, we focus on the effect of the $O_2$ addition rate on the spatial heterogeneity of oxidative aging. Figure 7 presents the static structure factor $S(q)$ calculated between scission sites. An upturn observed in the low-$q$ region of $S(q)$ indicates spatial heterogeneity [28]. A clear dependence of heterogeneity on $k_2$ was identified, consistent across both low and high $\alpha$ regions. As aging progresses, the heterogeneity diminishes. This phenomenon, as previously reported [14,15], is attributed to the accelerated relaxation and enhanced molecular chain mobility due to oxidative chain scissions. In low-$k_2$ cases, the retardation of H-abstraction reactions causes radical diffusion to predominate over localized reaction progression, thereby explaining the observed reduction in heterogeneity with decreasing of $O_2$ concentrations.

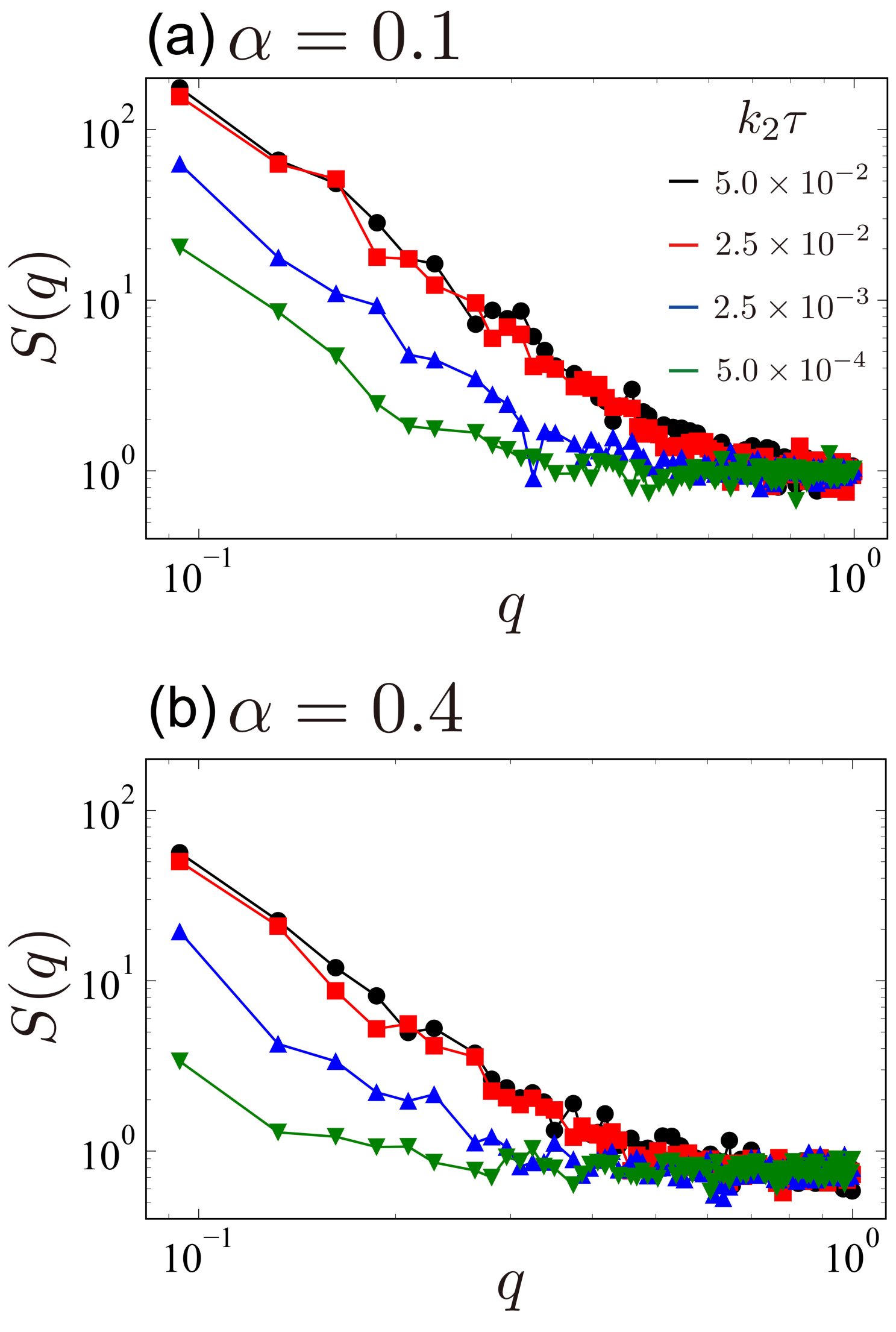

Figure 7 Static structure factor $S(q)$ between scission end beads for various $k_2$ cases: (a) $\alpha = 0.1$ (low conversion region), (b) $\alpha = 0.4$ (high conversion region).

Let us next consider another aspect reflecting the heterogeneity of oxidative aging: the spatial fluctuations of local conversion. As in our previous study [15], the spatial fluctuation of the local conversion is expressed using the number of remaining PH beads as follows.

$$\delta = \frac{\sqrt{\left\langle \left(N_{\mathrm{PH}}^{(i)}\right)^2 \right\rangle - \langle N_{\mathrm{PH}}^{(i)} \rangle^2}}{\left\langle N_{\mathrm{PH}}^{(i)} \right\rangle} \tag{13}$$

Here, we envisage the target simulation box divided into 20 subcells along each edge, resulting in 8000 small cubic regions. We focus on the fluctuations in the number of PH beads, $N_{\mathrm{PH}}^{(i)}$, present in the $i$-th domain. This fluctuation $\delta$ is equivalent to the fluctuation in the local conversion $\alpha_{\mathrm{local}}$. Figure 8 presents the spatial fluctuations of the local conversion, $\delta$, for different $k_2$ cases. The fluctuation $\delta$, reflecting aging heterogeneity, increases with higher $k_2$ conditions. This trend is consistent with the behavior observed in $S(q)$ calculated between scission sites.

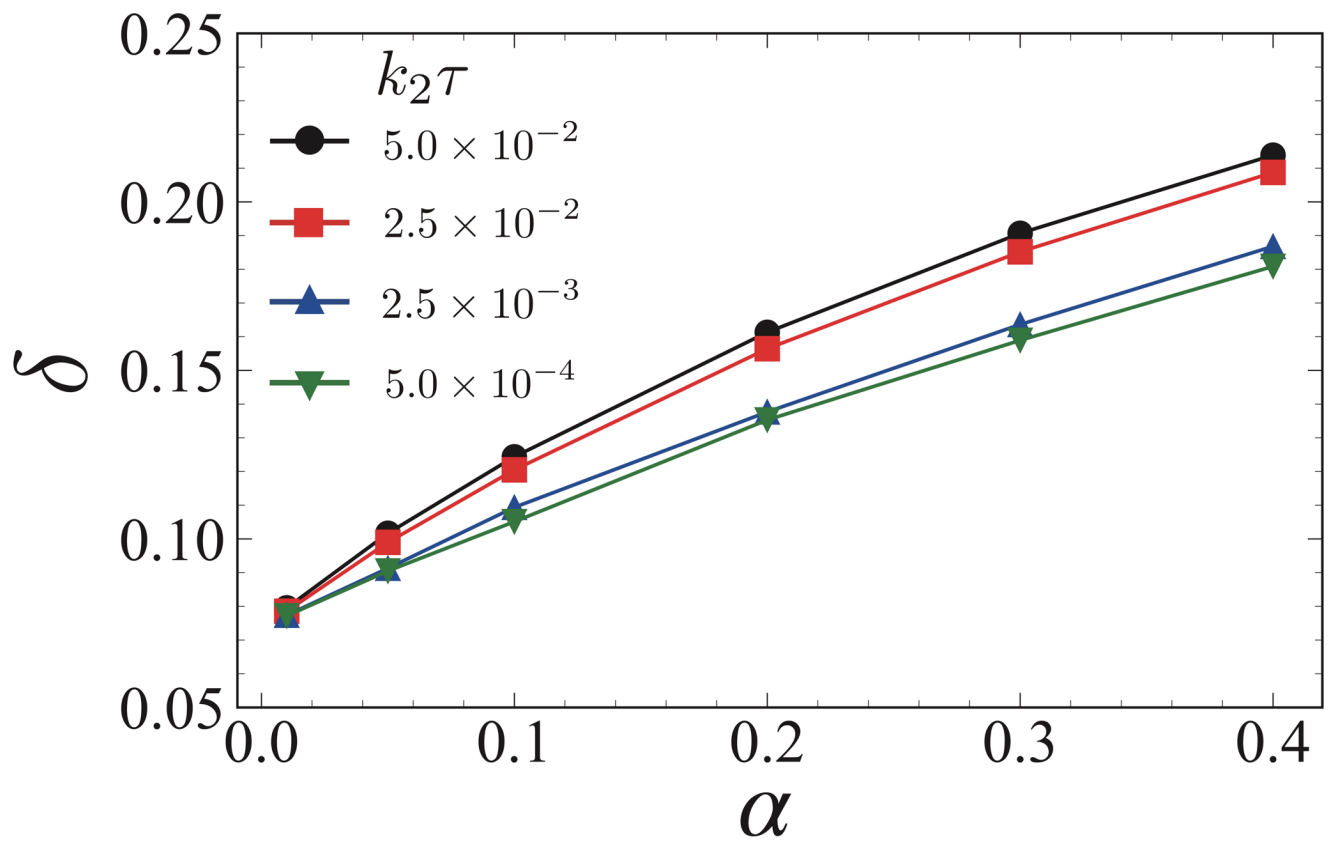

Figure 8 Spatial fluctuation $\delta$ for the number of PH beads, corresponding to the fluctuation of local conversion in oxidative aging.

The effect of $k_2$ extends to the chain length distribution, as shown in Figure 9. This figure illustrates the chain length distribution at $\alpha = 0.1$. When comparing the distributions at the same conversion level, it was observed that as $k_2$ decreases, the distribution broadens in the range of chain lengths between 5 and 30. This behavior is likely attributed to the enhanced impact of crosslinking between P· radicals under lower $O_2$ concentration conditions (see Figure 4).

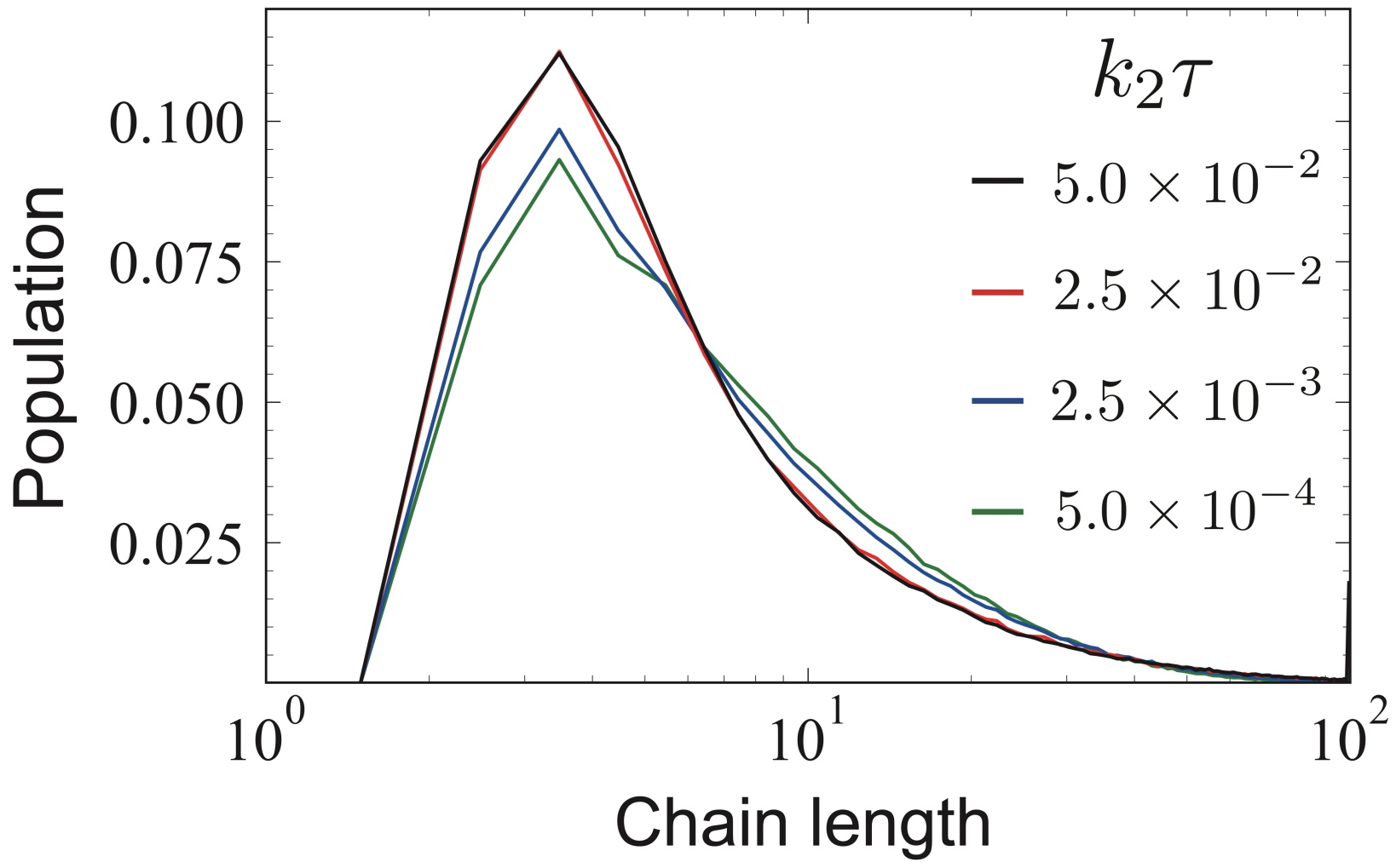

Figure 9 Distribution function of chain length distribution at conversion $\alpha = 0.1$ under different $k_2$ conditions.

Here, we summarize the influence of $O_2$ concentration on the behavior of oxidative aging. While this study varied the $O_2$ addition reaction rate $k_2$ in simulations to investigate the effect of $O_2$ concentration, similar control over $O_2$ addition reaction rate was achieved in experimental studies by adjusting $O_2$ partial pressure in autoclave environments [1–4]. Assuming that the onset of $[O_2]_{OER}$ corresponds to an $O_2$ partial pressure of 2 MPa [2], upper limit of $k_2\tau = 5.0 \times 10^{-2}$ corresponds to OER conditions, and the lower limit of $k_2\tau = 5.0 \times 10^{-4}$ corresponds to oxidative aging conditions in ambient air. In this study, the $O_2$ addition reaction converting P· to POO· slows down at low $k_2$ cases. Consequently, the effective rate of H-abstraction reactions, which are predominantly driven by POO· and govern the kinetics of oxidative aging, also decreases. This leads to an extended induction period and a slower reaction rate during the acceleration phase of oxidation reactions. Furthermore, the increased number of residual P· radicals enhances the rate of crosslinking reactions between P· radicals. The chain length distribution (i.e., molecular weight distribution) is also subtly affected, with the low $k_2$ cases exhibiting a broader chain length range for oxidized molecular chains. This broadening can be attributed to the influence of intermolecular crosslinking reactions. These findings do not contradict behaviors investigated in the field of polymer degradation through conventional reaction kinetic approaches [1,2,21]. The spatial heterogeneity of oxidative aging diminished as $k_2$ decreases, provided H-abstraction reaction rate ($k_3$) remains constant. Since the dynamics of macroradicals (P·, PO·, POO·) follow the dynamics of polymer segments, insights into the heterogeneity of oxidative aging can only be obtained through calculations of polymer dynamics. This highlights the unique contribution of such simulations in advancing our understanding of oxidative aging processes.

## 4. CONCLUSION

In this paper, we extended CGMD framework to perform oxidative aging simulations that reflect the effects of $O_2$ concentration via the reaction rate $k_2$ for the $O_2$ addition reaction. We conducted systematic simulations with varying $k_2$ to study $O_2$ concentration dependence for oxidative aging. The highest $k_2$ case ($k_2\tau = 5.0 \times 10^{-2}$) considered in this study corresponds to OER condition, while the lowest $k_2$ case ($k_2\tau = 5.0 \times 10^{-4}$) approximately corresponds to ambient air conditions. Reducing the $k_2$ value brings the aging system closer from the OER condition to the ambient air condition. From the aging simulations with varying $k_2$, the following key findings were revealed: (i) oxidative aging reactions are retarded, (ii) heterogeneity is mitigated, and (iii) the rate of crosslinking reactions increases. Except for the reaction retardation, these results could not have been elucidated without the simulation developed in this study. Expanding this framework to address photo-aging and aging in solid-state polymer is considered important future works in the field of polymer degradation. Furthermore, integrating this simulation technique as a mesoscale simulation module into a multiscale model capable of representing the DLO effect is also a promising future direction with significant importance for real industrial applications, and we plan to publish related work in the near future.

**Author Information**

**Corresponding Authors**

Takato Ishida - Department of Materials Physics, Nagoya University, Furo-cho, Chikusa, Nagoya 464-8601, Japan; orcid. Org/0000-0003-3919-2348; E-mail: ishida@mp.pse.nagoya-u.ac.jp

**Authors**

Kazuya Haremaki- Department of Materials Physics, Nagoya University, Furo-cho, Chikusa, Nagoya 464-8603, Japan; orcid. Org/0009-0007-8494-8008; E-mail: haremaki.kazuya.w2@s.mail.nagoya-u.ac.jp

Yusuke Koide - Department of Materials Physics, Nagoya University, Furo-cho, Chikusa, Nagoya 464-8603, Japan; orcid. Org/0000-0002-4843-6888; E-mail: koide.yusuke.k1@f.mail.nagoya-u.ac.jp

Takashi Uneyama - Department of Materials Physics, Nagoya University, Furo-cho, Chikusa, Nagoya 464-8603, Japan; orcid. Org/0000-0001-6607-537X; E-mail: uneyama@mp.pse.nagoya-u.ac.jp

Yuichi Masubuchi - Department of Materials Physics, Nagoya University, Furo-cho, Chikusa, Nagoya 464-8603, Japan; orcid. Org/0000-0002-1306-3823; E-mail: mas@mp.pse.nagoya-u.ac.jp

**Conflict of interest**

The author declares no competing interests.

**Acknowledgment**

This work was supported by JSPS KAKENHI Grant Numbers 24K20949, 22KJ1543, "Nagoya University High Performance Computing Research Project for Joint Computational Science" in Japan, CCI holdings Co., Ltd., Mayekawa Houonkai Foundation, Suzuki Foundation, Yazaki Memorial Foundation for Science and Technology, Fujimori Science and Technology Foundation, Fuji Seal

Foundation, The Naito Reserch Grant and Suga Weathering Technology Foundation

## Supporting information

# Coarse-Grained Molecular Dynamics Simulations for Oxidative Aging of Polymers under Various $O_2$ Concentrations

*Takato Ishida*[*], *Kazuya Haremaki, Yusuke Koide, Takashi Uneyama, Yuichi Masubuchi*

Department of Materials Physics, Nagoya University, Furo-cho, Chikusa, Nagoya 464-8603, Japan

Tel: +81-52-789-4202, Mobile:+81-90-1145-4372

*E-mail address:* ishida@mp.pse.nagoya-u.ac.jp

TABLE OF CONTENTS

**1. Normalized time evolution of chemical kinetic behaviors**

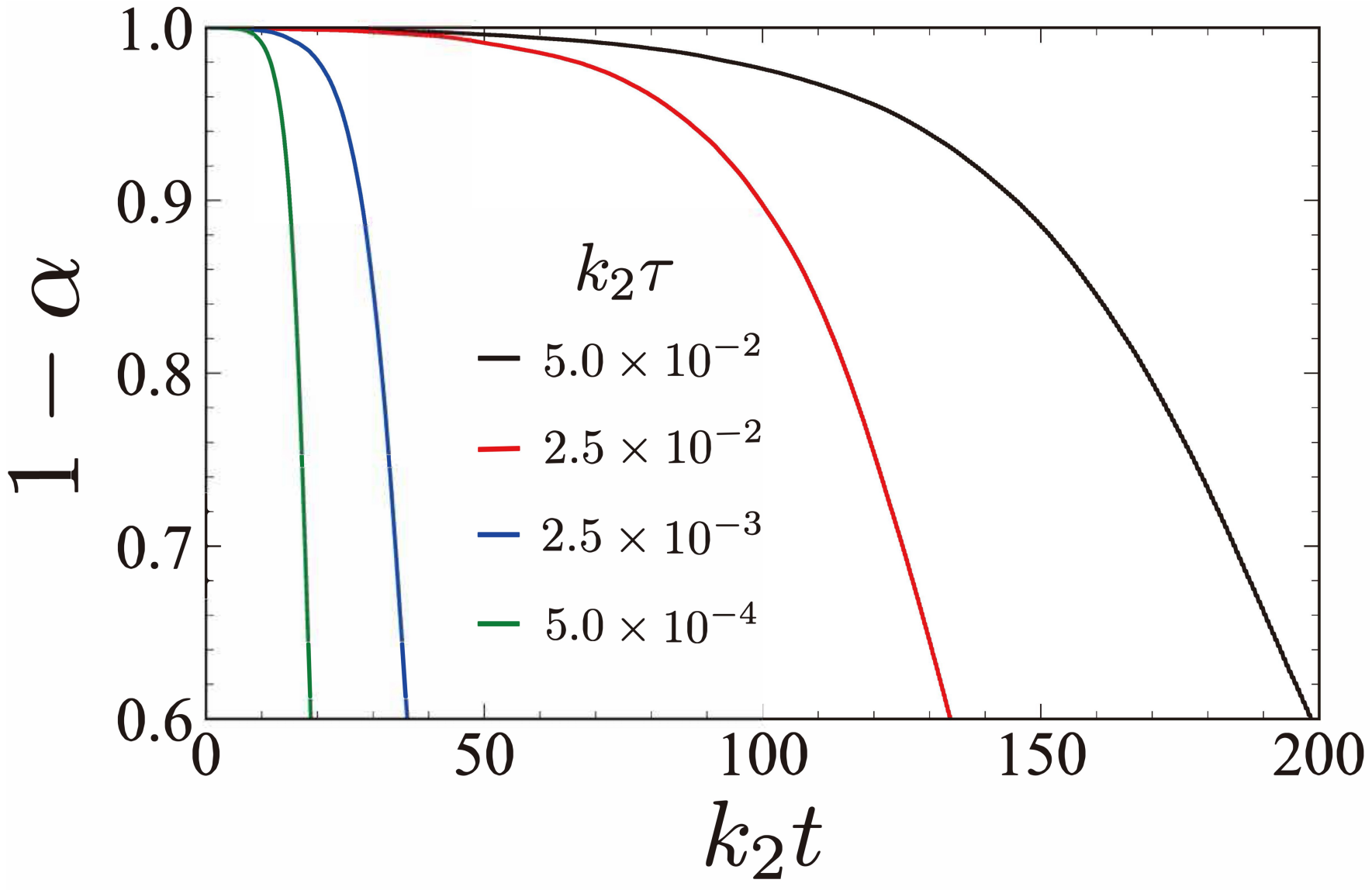

Figure S1 Relationship between conversion ($\alpha$) and the time normalized by the $O_2$ addition reaction rate ($k_2t$) at various $k_2$ cases.

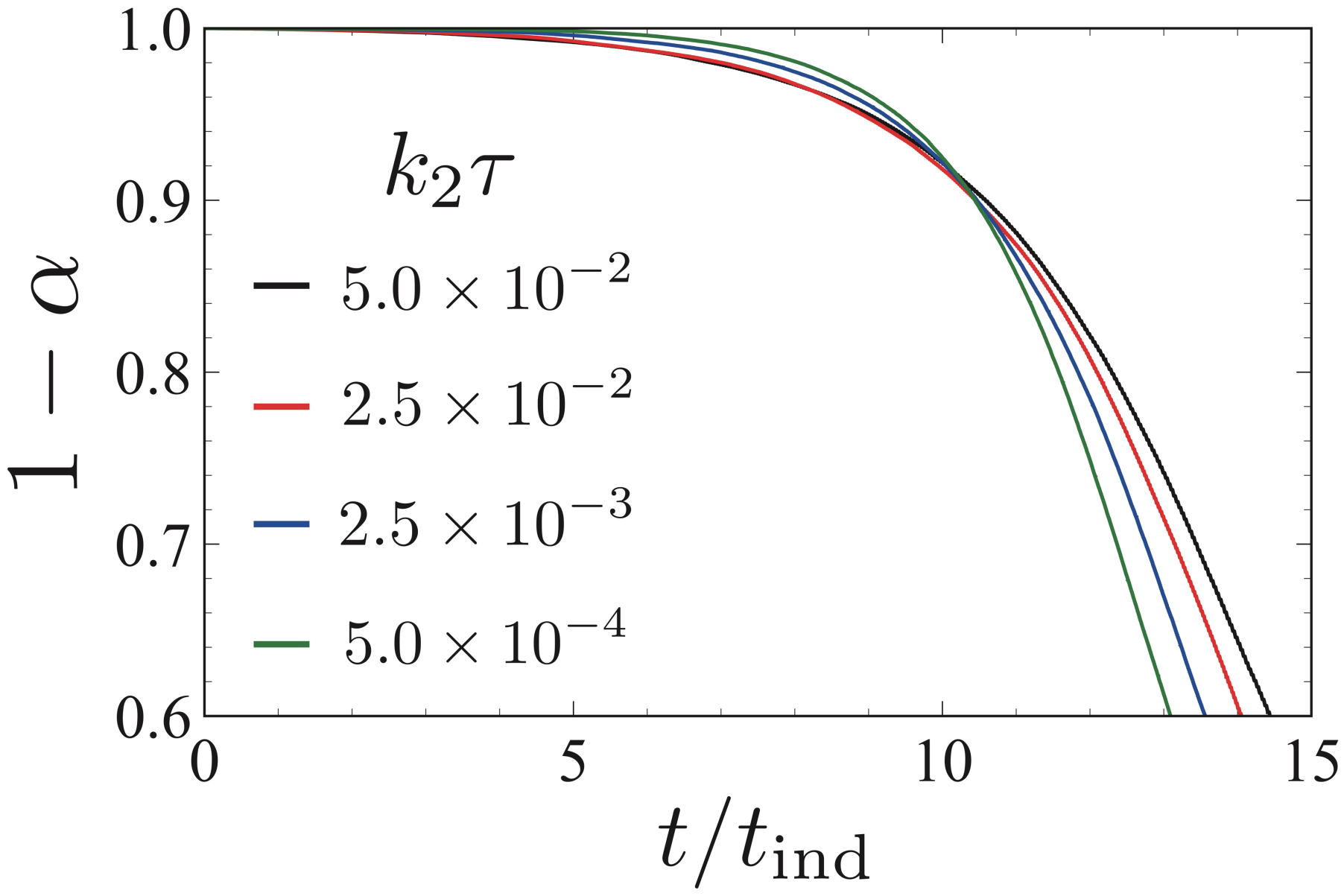

Figure S2 Relationship between conversion ($\alpha$) and dimensionless time ($t/t_{\mathrm{ind}}$) at various $k_2$ cases.

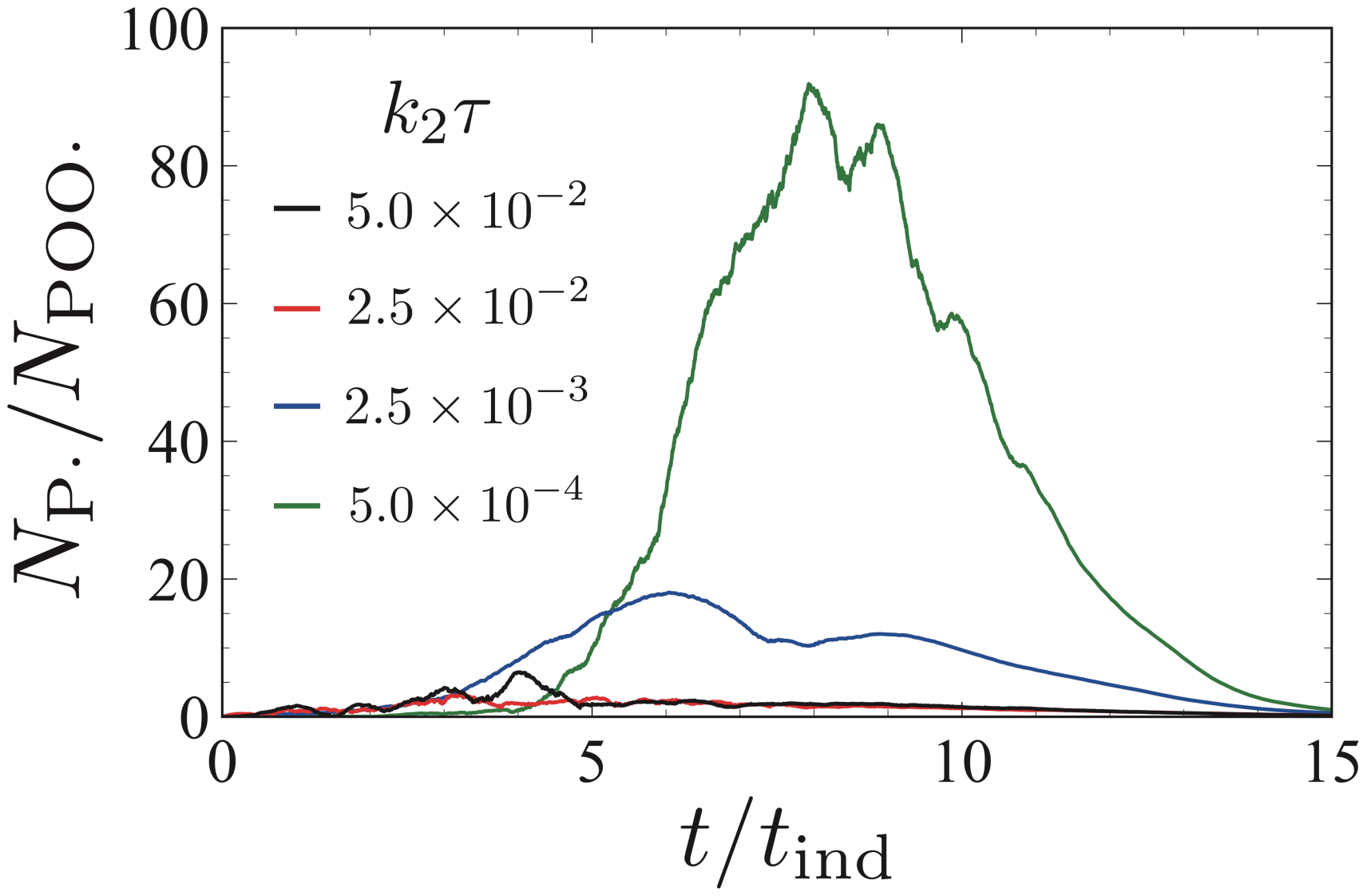

Figure S3 Time evolution of the ratio of P· to POO· radicals ($N_{P\cdot}/N_{POO\cdot}$) as a function of $t/t_{\text{ind}}$ at various $k_2$ cases.

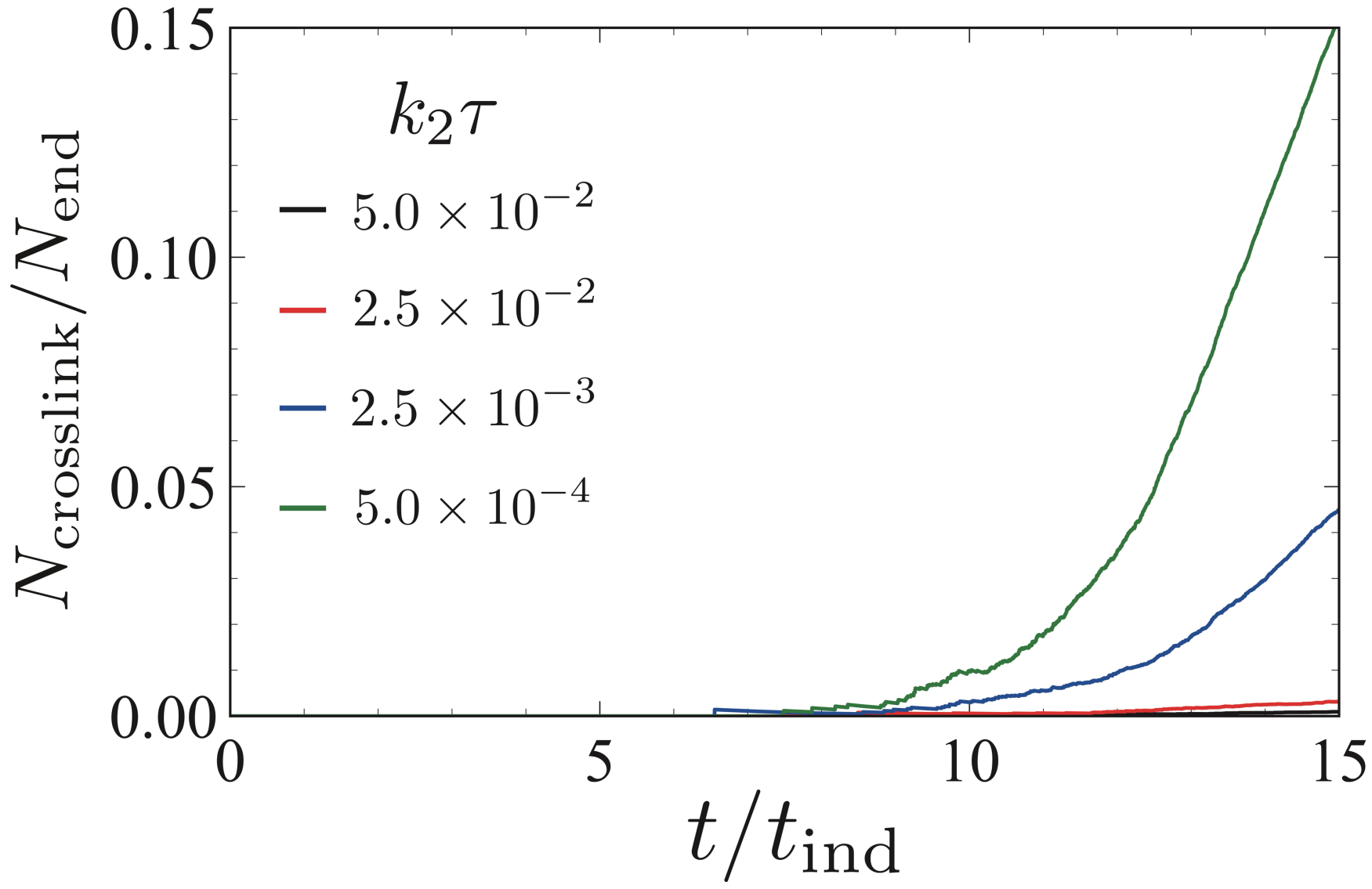

Figure S4 Time evolution of the ratio of crosslink to scission end counts ($N_{\text{crosslink}}/N_{\text{end}}$) as a function of $t/t_{\text{ind}}$ at various $k_2$ cases.